\documentclass[aps,prl,10pt,twocolumn,superscriptaddress,nofootinbib]{revtex4-2}

\usepackage{graphicx}
\usepackage{bm}
\usepackage{color}
\usepackage[colorlinks=true,pdfstartview=FitV,linkcolor=blue,citecolor=blue,urlcolor=blue]{hyperref}
\usepackage[dvipsnames]{xcolor}
\usepackage{amsmath, amsfonts}
\usepackage{booktabs}
\usepackage[caption=false]{subfig} 
\usepackage[normalem]{ulem}

\enlargethispage{\baselineskip}
\everymath{\displaystyle}

\begin{document}

\title{Constraining exotic high-\texorpdfstring{$z$}{z} reionization histories with Gaussian processes and the Cosmic Microwave Background}

\newcommand{\SMPS}{\affiliation{School of Mathematical and Physical Sciences, University of Sheffield,
Hounsfield Road, Sheffield S3 7RH, United Kingdom}}
\newcommand{\KCL}{\affiliation{Theoretical Particle Physics and Cosmology, King’s College London, Strand, London, WC2R 2LS, United Kingdom}}
\newcommand{\TDLI}{\affiliation{Tsung-Dao Lee Institute \& School of Physics and Astronomy, Shanghai Jiao Tong University, Shanghai 201210, China}}

\author{Hanyu Cheng}
\email{chenghanyu@sjtu.edu.cn}
\TDLI \SMPS
\thanks{HC and ZY contributed equally to this work \newline }

\author{Ziwen Yin}
\email{ziwenyin@sjtu.edu.cn}
\TDLI \KCL
\thanks{HC and ZY contributed equally to this work \newline }

\author{Eleonora Di Valentino}
\email{e.divalentino@sheffield.ac.uk}
\SMPS

\author{David J. E.\ Marsh}
\email{david.j.marsh@kcl.ac.uk}
\KCL

\author{Luca Visinelli}
\email{lvisinelli@unisa.it}
\affiliation{Dipartimento di Fisica ``E.R.\ Caianiello'', Universit\`a degli Studi di Salerno,\\ Via Giovanni Paolo II, 132 - 84084 Fisciano (SA), Italy}
\affiliation{Istituto Nazionale di Fisica Nucleare - Gruppo Collegato di Salerno - Sezione di Napoli,\\ Via Giovanni Paolo II, 132 - 84084 Fisciano (SA), Italy}

\begin{abstract}
The large-angle polarization anisotropies in the Cosmic Microwave Background (CMB) arise from Thomson scattering of CMB photons off free electrons in the post-recombination Universe. In the standard $\Lambda$ cold dark matter cosmological model, the free electron density increases at redshifts $z \lesssim 10$ as the first stars form, reionizing the intergalactic medium. We use \emph{Gaussian processes} to perform a model-independent reconstruction of the cosmic reionization history constrained by \textit{Planck} CMB data. Our approach recovers the standard reionization at $z \lesssim 10$ and places stringent limits on any additional high-$z$ reionization. From this reconstruction, we define a new derived parameter, the high-redshift contribution to the CMB optical depth, $\tau_{\mathrm{highz}}$, whose posterior distribution provides robust constraints on exotic energy injection scenarios. We demonstrate this for decaying dark matter with particle masses in the range $\mathcal{O}(1\,\text{MeV})$. A companion paper applies this framework to multi-axion models. All data and code are publicly available at: \href{https://github.com/Cheng-Hanyu/CLASS_reio_gpr}{github.com/Cheng-Hanyu/CLASS\_reio\_gpr}.
\end{abstract}

\date{\today}
\maketitle


\textbf{\textit{Introduction ---}} Cosmological observations provide a detailed reconstruction of the large-scale evolution of the Universe and offer a unique window into the fundamental particles and interactions that shaped its earliest moments. Measurements of Big Bang Nucleosynthesis (BBN)~\cite{Fields:2019pfx} and the Cosmic Microwave Background (CMB)~\cite{Planck:2018nkj, Planck:2018vyg} are particularly sensitive to light relics, such as the Standard Model (SM) neutrinos, and to the primordial baryon density. These probes also reveal the presence of dark matter (DM)~\cite{Bertone:2004pz, Feng:2010gw, Cirelli:2024ssz} and dark energy (DE)~\cite{SupernovaSearchTeam:1998fmf, SupernovaCosmologyProject:1998vns}, indicating new physics beyond the SM. The dark sector may further include light or relativistic species contributing to the effective number of neutrino-like degrees of freedom~\cite{Baumann:2016wac}, leaving measurable imprints on cosmological observables.

Combining cosmological probes across different redshifts provides a powerful means to break parameter degeneracies and tighten bounds on new physics~\cite{Weinberg:2013agg, SimonsObservatory:2018koc}. Joint analyses of early-Universe data, such as CMB anisotropies~\cite{Planck:2018nkj, Planck:2018vyg, ACT:2020gnv, ACT:2025fju, SPT-3G:2024atg}, with low-redshift observables including large-scale structure~\cite{DES:2024jxu, Wright:2025xka}, baryon acoustic oscillations (BAO)~\cite{eBOSS:2020yzd, DESI:2025zgx}, and supernovae~\cite{Brout:2022vxf, DES:2024jxu, Rubin:2023ovl} allow correlated parameters to be disentangled and enhance sensitivity to extensions of the standard cosmological model ($\Lambda$CDM). Such analyses have enabled stringent tests of time-varying DE~\cite{Copeland:2006wr,Cooray:1999da,Efstathiou:1999tm,Chevallier:2000qy,Linder:2002et,Wetterich:2004pv,Feng:2004ff,Hannestad:2004cb,Xia:2004rw,Gong:2005de,Jassal:2005qc,Nesseris:2005ur,Liu:2008vy,Barboza:2008rh,Barboza:2009ks,Ma:2011nc,Sendra:2011pt,Feng:2011zzo,Barboza:2011gd,DeFelice:2012vd,Feng:2012gf,Wei:2013jya,Magana:2014voa,Akarsu:2015yea,Pan:2016jli,DiValentino:2016hlg,Nunes:2016plz,Nunes:2016drj,Magana:2017usz,Yang:2017alx,Pan:2017zoh,Panotopoulos:2018sso,Yang:2018qmz,Jaime:2018ftn,Das:2017gjj,Yang:2018prh,Li:2019yem,Yang:2019jwn,Pan:2019hac,Tamayo:2019gqj,Pan:2019brc,DiValentino:2020naf,Rezaei:2020mrj,Perkovic:2020mph,Banihashemi:2020wtb,Jaber-Bravo:2019nrk,Benaoum:2020qsi,Yang:2021eud,Jaber:2021hho,Alestas:2021luu,Yang:2022klj,Escudero:2022rbq,Castillo-Santos:2022yoi,Yang:2022kho,Dahmani:2023bsb,Escamilla:2023oce,Rezaei:2023xkj,Adil:2023exv,LozanoTorres:2024tnt,Singh:2023ryd,Rezaei:2024vtg,Reyhani:2024cnr,DESI:2024mwx,Cortes:2024lgw,Shlivko:2024llw,Luongo:2024fww,Yin:2024hba,Gialamas:2024lyw,Dinda:2024kjf,Najafi:2024qzm,Wang:2024dka,Ye:2024ywg,Tada:2024znt,Carloni:2024zpl,Park:2024pew,DESI:2024kob,Ramadan:2024kmn,Notari:2024rti,Orchard:2024bve,Hernandez-Almada:2024ost,Malekjani:2024bgi,Giare:2024gpk,Reboucas:2024smm,Giare:2024ocw,Menci:2024hop,Li:2024qus,Li:2024hrv,Notari:2024zmi,Gao:2024ily,Fikri:2024klc,Jiang:2024xnu,Zheng:2024qzi,Gomez-Valent:2024ejh,RoyChoudhury:2024wri,Lewis:2024cqj,Wolf:2025jlc,Shajib:2025tpd,Giare:2025pzu,Chaussidon:2025npr,Kessler:2025kju,CosmoVerse:2025txj,Pang:2025lvh,RoyChoudhury:2025dhe,Scherer:2025esj,Specogna:2025guo,Cheng:2025lod,Cheng:2025hug}, modified gravity~\cite{Clifton:2011jh, Nojiri:2006ri,Nojiri:2017ncd, Koyama:2015vza,Carroll:2004de, Hu:2007pj,CANTATA:2021asi, Ishak:2024jhs, Taule:2024bot, Chudaykin:2024gol, Specogna:2023nkq, Specogna:2024euz}, and non-standard thermal histories~\cite{Bernal:2016gxb, Grin:2007yg, Konings:2024zvz, Waldstein:2016blt, Visinelli:2009kt}.

\begin{figure}[htb]
    \centering
    \includegraphics[width=\linewidth]{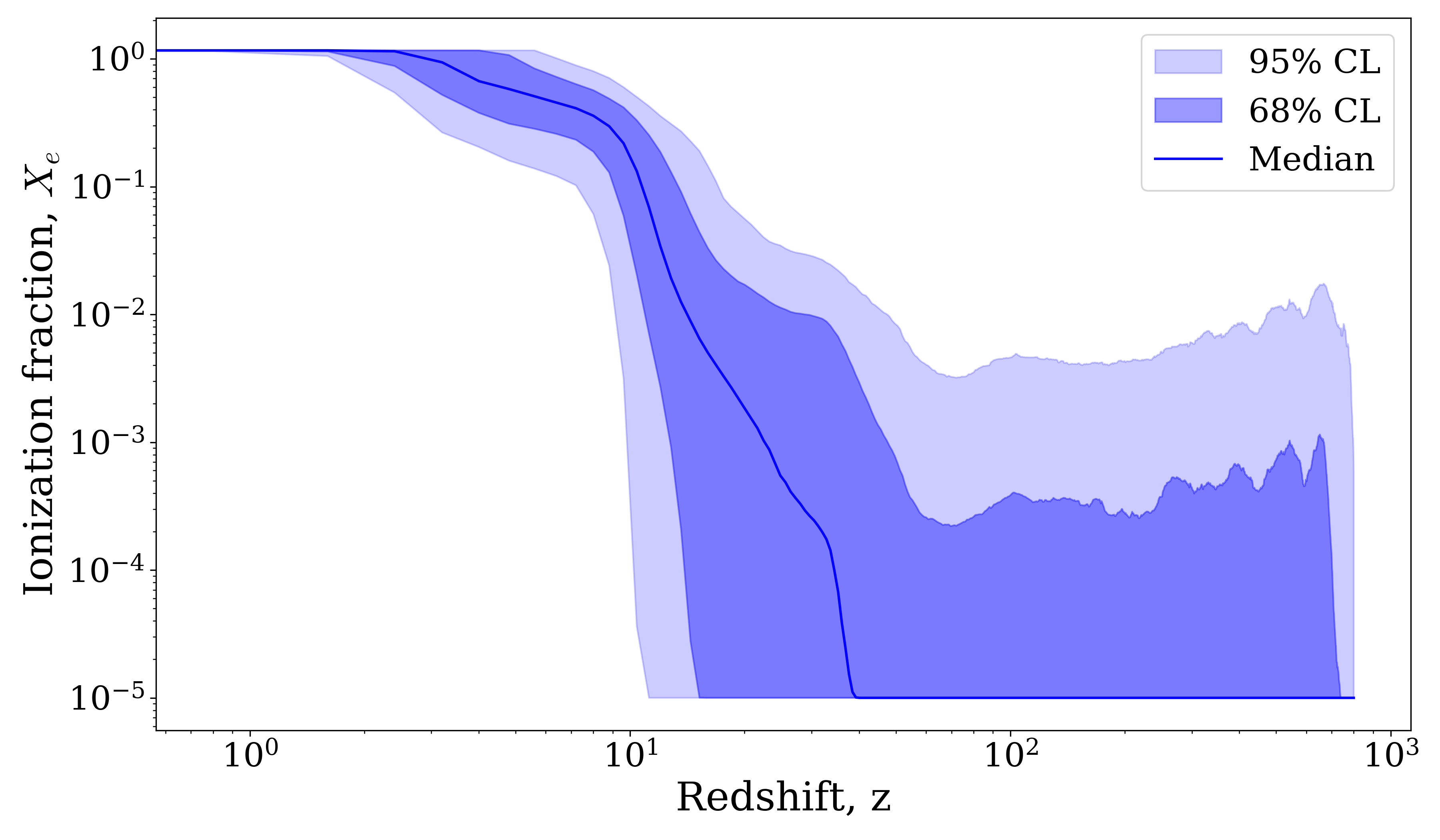}
    \caption{Model independent reconstruction of the free electron fraction, $X_e(z)$, from \textit{Planck} large angle E-mode polarization data, using the Gaussian process method. All other cosmological parameters are fixed to their best-fit values from the \textit{Planck} 2018 TT,TE,EE+lowE analysis~\cite{Planck:2018vyg}. The shaded regions denote the 68\% (blue) and 95\% (light blue) confidence level (CL), and the solid line marks the median reconstruction.}
    \label{fig:Reconstruction_X_e}
\end{figure} 

An additional cosmological probe, less exploited than CMB primary anisotropies or large-scale structure, is cosmic reionization. This epoch contains key information on the thermal and ionization history of the Universe and offers sensitivity to new physics, particularly from heavy or unstable relics. The ionization fraction,
\begin{equation}
    X_e(z) \equiv n_e(z)/n_{\mathrm{H}}(z),,
\end{equation}
denotes the number of free electrons per hydrogen nucleus at redshift $z$, with $n_e(z)$ and $n_{\mathrm{H}}(z)$ the respective number densities. This quantity governs the thermal and radiative history of the Universe, especially during recombination and reionization. During recombination, at $z_{\rm rec} \sim 1100$, electrons and protons formed neutral hydrogen, evolving from Saha equilibrium to out-of-equilibrium multi-level atomic processes~\cite{Peebles:1968ja, Seager:1999bc}. Variations in $X_e(z)$ after recombination left imprints on the CMB anisotropies, making accurate modeling of this quantity essential for cosmological inference~\cite{Planck:2018nkj}.

During reionization, astrophysical sources such as the first stars, galaxies, and accreting black holes (BHs) reionized the intergalactic medium (IGM), raising $X_e$ from near zero to unity by $z \sim 6$~\cite{Fan:2006dp, Robertson:2015uda}. The timing and duration of reionization affect large-scale CMB polarization and the integrated optical depth $\tau_{\mathrm{reio}}$, constraining models of early structure formation and ionizing radiation. Many extensions of $\Lambda$CDM alter this history, including DM annihilation~\cite{Galli:2009zc, 2011PhRvD..84b7302G, 2012PhRvD..85d3522F, Galli:2013dna} and decay~\cite{Diamanti:2013bia, Lopez-Honorez:2013cua, Poulin:2016anj, Lopez-Honorez:2016sur, Cheung:2018vww}, axions~\cite{Marsh:2015xka, Escudero:2023vgv}, feebly interacting relics~\cite{Covi:1997my, Feng:2003uy, Hall:2009bx}, strongly interacting relics~\cite{Escudero:2018thh}, warm DM~\cite{Lopez-Honorez:2017csg}, and primordial BHs~\cite{Poulin:2017bwe, Mena:2019nhm, Agius:2024ecw}.

In this \textit{Letter}, we derive model-independent constraints on the ionization fraction $X_e(z)$ using Gaussian Process (GP) regression of \textit{Planck} CMB polarization data, allowing arbitrary departures from $\Lambda$CDM at $z < z_{\rm rec}$. The resulting reconstruction is shown in Fig.~\ref{fig:Reconstruction_X_e}. We then compress the constrained $X_e(z)$ realizations into two effective parameters describing the low- and high-redshift contributions to the optical depth and derive their posteriors. Using a decaying DM scenario as a case study, we demonstrate that the posterior on the high-redshift optical-depth parameter $\tau_{\rm high}$ provides a robust and efficient probe of new physics beyond $\Lambda$CDM. A companion Paper~II~\cite{Yin:2025amn} applies this framework to constrain the reheating temperature and explore its relation to the topology of extra dimensions in string compactifications~\cite{Gendler:2023kjt}. The analysis code is publicly available at \href{https://github.com/Cheng-Hanyu/CLASS_reio_gpr}{github.com/Cheng-Hanyu/CLASS\_reio\_gpr}, while the companion code can be found at \href{https://github.com/ZiwenYin/Reionization-with-multi-axions-decay}{github.com/ZiwenYin/Reionization-with-multi-axions-decay}.

\textbf{\textit{Methodology and Data ---}} We perform Bayesian inference using the \texttt{Cobaya} framework~\cite{Torrado:2020dgo}, with a Markov Chain Monte Carlo (MCMC) sampler tailored for cosmological analyses. This is combined with a modified version of the Boltzmann code Cosmic Linear Anisotropy Solving System (\texttt{CLASS})~\cite{Blas:2011rf}, extended to support multiple reionization schemes, including the \texttt{reio\_gpr\_tanh} case introduced in the Supplemental Material. This extension enables data-driven reconstructions of the ionization history. Convergence is monitored using the Gelman--Rubin statistics $R - 1 < 0.1$, and posterior distributions are analyzed using \texttt{getdist}~\cite{Lewis:2019xzd}.

We employ the low-$\ell$ EE-only SimAll likelihood from the \textit{Planck} 2018 release~\cite{Planck:2018nkj, Planck:2019nip}, which provides the cleanest probe of reionization. Large-scale polarization anisotropies arise directly from Thomson scattering during reionization, making the optical depth
\begin{equation}
    \tau_{\mathrm{reio}} = \int^{z_{\mathrm{max}}}_0 \sigma_T\,n_e(z)\,\frac{\mathrm{d}z}{(1+z)H(z)}\,,
    \label{eq:tau_reio}
\end{equation}
the key parameter controlling the low-$\ell$ EE amplitude. We set $z_{\mathrm{max}} = 800$ to allow for early ionization scenarios while remaining below the recombination epoch, isolating post-recombination physics.

High-redshift contributions to $\tau_{\mathrm{reio}}$ leave distinct imprints at angular scales set by the horizon size when the free electron fraction increases significantly. Low-$\ell$ polarization data alone provides robust constraints on $X_e(z < z_{\rm rec})$, with high-$\ell$ modes contributing negligibly. The SimAll likelihood benefits from end-to-end simulation-based noise and systematics, making it the most robust dataset for this analysis.

The GP reconstruction introduces a large number of additional parameters. To ensure convergence with feasible computational resources, we fix the standard $\Lambda$CDM parameters to their \textit{Planck} 2018 TT,TE,EE+lowE best-fit values, summarized in Table~\ref{tab:prior_table}. Our method directly reconstructs the free electron fraction history, $X_e(z)$, the fundamental quantity governing the CMB E-mode polarization spectrum. This approach avoids imposing specific reionization parameterizations~\cite{Ilic:2025idl}, providing a more direct and physically transparent link to the data. The optical depth is then obtained as a derived quantity, computed internally by the modified \texttt{CLASS} solver.

\begin{table}[t!]
\centering
\begin{tabular}{lcc}
\hline
Parameter & Fixed Value \\
\hline
$\Omega_\mathrm{b}h^2$ & 0.02236 \\
$\Omega_\mathrm{c}h^2$ & 0.1202 \\
$A_\mathrm{s}$ & $2.101 \times 10^{-9}$ \\
$n_\mathrm{s}$ & 0.9649 \\
$H_0$ [km s$^{-1}$ Mpc$^{-1}$] & 67.27 \\
\hline
\end{tabular}
\caption{Fixed cosmological parameters used as priors in this analysis.}
\label{tab:prior_table}
\end{table}

To implement the new GP-driven reionization scheme, we introduce additional parameters and routines in the \texttt{CLASS} reionization module~\cite{Blas:2011rf} (see Supplemental Material). We adopt a non-uniform interpolation strategy for reconstructing the ionization fraction history $X_e(z)$,\footnote{For comparison, we also provide an alternative reconstruction using uniformly spaced interpolation bins in the Supplemental Material.} with finer binning at low redshift and coarser sampling at high redshift. This choice reflects the rapid variation of $X_e(z)$ at low redshift, requiring higher resolution to capture the transition, while at high redshift the ionization fraction remains close to neutral and evolves slowly, allowing for wider bins without loss of accuracy. The non-uniform binning enables reconstructions up to $z = 800$ while keeping computations tractable, which is essential for capturing exotic early energy injection and modeling the full reionization history. The GP-based scheme is implemented in the \texttt{gp\_model} model, with parameter priors summarized in Table~\ref{tab:priors}.

\begin{table}[htbp]
\centering
\begin{tabular}{lcc}
\toprule
\textbf{Parameter} & \textbf{Prior Range} & \textbf{Reference Value} \\
\midrule
\textbf{\texttt{gpr\_n\_low}} & $[2, 5]$ & 3 \\
\textbf{\texttt{gpr\_n\_high}} & $[10, 20]$ & 15 \\
\textbf{\texttt{z\_transition}} & $[10.0, 50.0]$ & 20.0 \\
\textbf{\texttt{gpr\_reio\_step\_sharpness}} & $[0.1, 1.0]$ & 0.5 \\
\textbf{\texttt{sigma\_f}} & $[0.1, 1.0]$ & 0.5 \\
\textbf{\texttt{l\_gpr}} & $[2.0, 30.0]$ & 5.0 \\
\textbf{\texttt{z\_min}} & $[4.0, 8.0]$ & 6.0 \\
\textbf{\texttt{z\_max}} & $[700.0, 800.0]$ & 750.0 \\
\bottomrule
\end{tabular}
\caption{Prior for the parameters introduced in the modified \texttt{CLASS} reionization module used in the GP interpolation method.}
\label{tab:priors}
\end{table}

To isolate the contribution of different epochs to the total optical depth, we decompose $\tau_{\mathrm{reio}}$ into low- a high-redshift components, $\tau_{\mathrm{lowz}}$ and $\tau_{\mathrm{highz}}$, respectively. This separation distinguishes between standard astrophysical reionization and potential exotic mechanisms that ionize the Universe at earlier times. Following Eq.~\eqref{eq:tau_reio}, we define
\begin{equation}
\label{eq:tau_lowhighz}
\begin{split}
    \tau_{\mathrm{lowz}} &= \int^{z_c}_0 \sigma_T\,n_e(z)\,\frac{\mathrm{d}z}{(1+z)H(z)}\,,\\
    \tau_{\mathrm{highz}} &= \int^{z_{\mathrm{max}}}_{z_c} \sigma_T\,n_e(z)\,\frac{\mathrm{d}z}{(1+z)H(z)}\,,
\end{split}
\end{equation}
in terms of a critical redshift $z_c$, so that $\tau_{\mathrm{reio}} = \tau_{\mathrm{lowz}} + \tau_{\mathrm{highz}}$.

While current CMB observations primarily constrain the total optical depth, this decomposition provides physical insight into which redshift ranges dominate the signal. The quantity $\tau_{\mathrm{lowz}}$ corresponds to standard reionization driven by the first stars and galaxies, which becomes efficient at $z \sim 20$--$30$ in $\Lambda$CDM cosmology. In contrast, $\tau_{\mathrm{highz}}$ captures any early ionization induced by exotic processes such as DM decay, BH accretion, or other energy injection mechanisms. In standard scenarios, $\tau_{\mathrm{highz}}$ is expected to be negligible: a statistically significant detection would indicate new physics.

\begin{figure*}[bht]
    \centering
    \includegraphics[width=\linewidth]{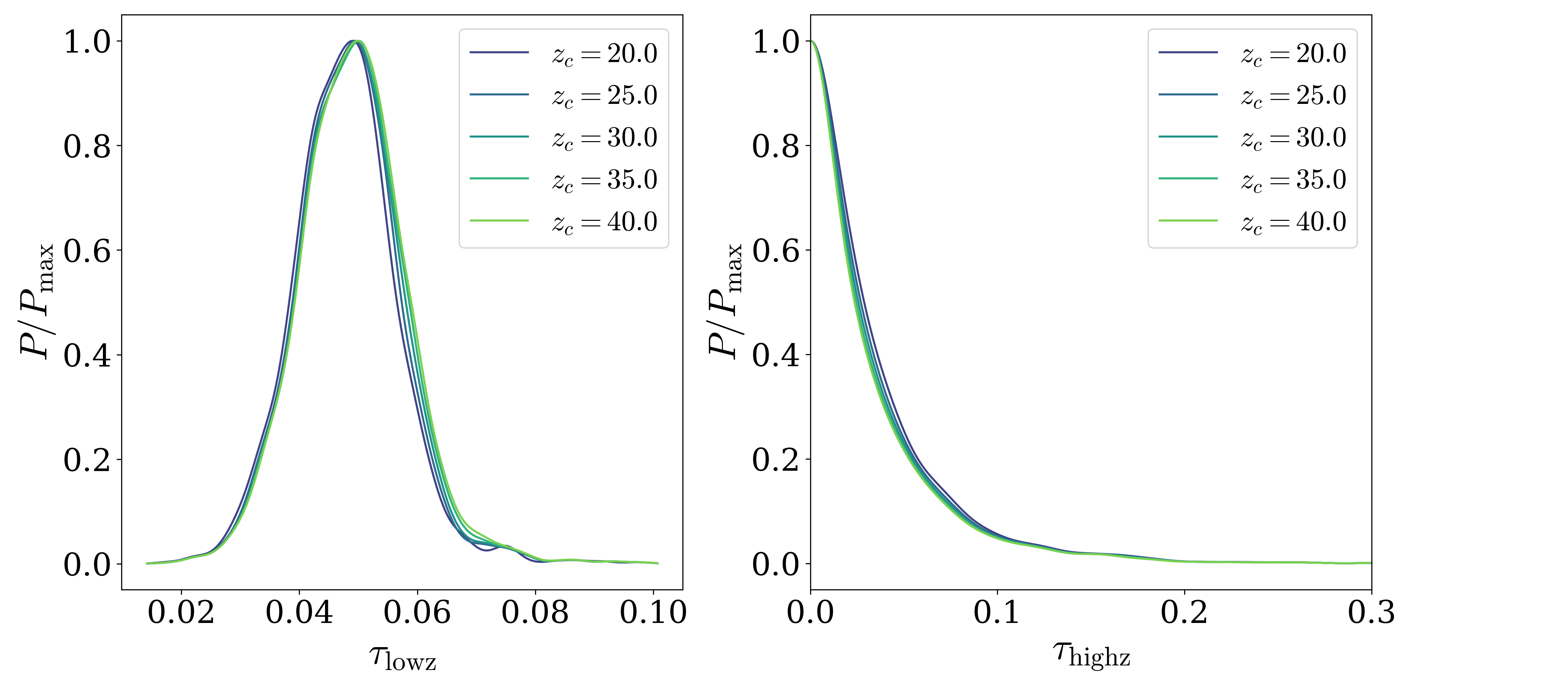}
\caption{Normalized posterior distributions $P/P_{\mathrm{max}}$ for the low-redshift (left) and high-redshift (right) contributions to the optical depths, shown for different critical redshifts $z_{\rm c}$ separating the two regimes, as listed in Table~\ref{tab:tau_constraints}. Results are based on \textit{Planck} low-$\ell$ EE data alone.}
    \label{fig:tau_zc}
\end{figure*}

\textbf{\textit{Results ---}} Table~\ref{tab:tau_constraints} summarizes the inferred constraints on the low- and high-redshift contributions to the total optical depth, $\tau_{\rm lowz}$ and $\tau_{\mathrm{highz}}$, as functions of the critical redshift $z_{\rm c}$. The inferred $\tau_{\rm lowz}$ values are remarkably stable across all $z_{\rm c}$, with $\tau_{\rm lowz} \approx 0.048$ and 68\%~CL uncertainties of $\sim0.008$. This robustness indicates that current CMB polarization data constrain the low-redshift contribution to reionization independently of the transition redshift. In contrast, $\tau_{\mathrm{highz}}$ is limited only by an upper bound, $\tau_{\mathrm{highz}} < 0.111$--$0.112$ (95\%~CL), nearly independent of $z_{\rm c}$. As $z_{\rm c}$ increases, $\tau_{\rm lowz}$ slightly increases while $\tau_{\mathrm{highz}}$ decreases, consistent with shifting a small fraction of the ionization history into the low-$z$ range. The small amplitude of this effect confirms the limited sensitivity of current data to separating early and late ionization contributions.

\begin{table}[htbp]
    \centering
    \begin{tabular}{c|cc|cc}
        \hline\hline
        $z_{\rm c}$ & \multicolumn{2}{c}{68\% CL} & \multicolumn{2}{c}{95\% CL} \\
        \cline{2-5}
        & $\tau_{\rm lowz}$ & $\tau_{\mathrm{highz}}$ & $\tau_{\rm lowz}$ & $\tau_{\mathrm{highz}}$ \\
        \hline
        20.0 & $0.0471^{+0.0076}_{-0.0082}$ & $<0.038$ & $0.047^{+0.016}_{-0.017}$ & $<0.112$ \\
        25.0 & $0.0477^{+0.0076}_{-0.0082}$ & $<0.038$ & $0.048^{+0.016}_{-0.017}$ & $<0.112$ \\
        30.0 & $0.0480^{+0.0078}_{-0.0082}$ & $<0.037$ & $0.048^{+0.016}_{-0.017}$ & $<0.111$ \\
        35.0 & $0.0483^{+0.0079}_{-0.0083}$ & $<0.037$ & $0.048^{+0.016}_{-0.017}$ & $<0.111$ \\
        40.0 & $0.0486^{+0.0080}_{-0.0084}$ & $<0.037$ & $0.049^{+0.016}_{-0.018}$ & $<0.111$ \\
        \hline\hline
    \end{tabular}
\caption{Constraints on the low-redshift $\tau_{\rm lowz}$ and high-redshift $\tau_{\mathrm{highz}}$ contributions to the optical depths for different choices of the critical redshift $z_{\rm c}$. For each parameter, we report the mean value and 68\% CL, or the 68\% and 95\% CL upper limits.}
\label{tab:tau_constraints}
\end{table}

The high-redshift optical depth $\tau_{\mathrm{highz}}$ probes the earliest ionizing sources, such as metal-free Population~III (PopIII) stars forming at $z \gtrsim 15\text{--}30$, whose energetic emission and associated core-collapse supernovae could partially ionize the IGM. Table~\ref{tab:tau_constraints} places bounds on such early ionization, including possible PopIII contributions~\cite{Hartwig:2022lon}. Additional sources from accreting BHs, X-ray binaries, or exotic processes, may also contribute to $\tau_{\mathrm{highz}}$. The James Webb Space Telescope (JWST) observations have begun probing galaxies at $z \gtrsim 10$~\cite{Gardner:2006ky}, but the absence of PopIII signatures leaves room for non-standard ionizing sources~\cite{2022ApJ...940L..55F, 2022ApJ...940L..14N, Iocco:2024rez}. In this context, our GP-based reconstruction of $X_e(z)$ from CMB data alone provides robust constraints on the allowed early ionization budget. Combining these results with JWST constraints on high-redshift galaxy populations will help determine if standard astrophysical sources are sufficient to account for the allowed $\tau_{\mathrm{highz}}$, or if additional, possibly exotic, contributions are required.

Figure~\ref{fig:tau_zc} shows the normalized posterior distributions for $\tau_{\rm lowz}$ and $\tau_{\mathrm{highz}}$ obtained from \textit{Planck} low-$\ell$ EE polarization data. The posteriors for $\tau_{\rm lowz}$ are sharply peaked and nearly invariant with $z_{\rm c}$, confirming the robustness of the low-redshift constraint. By contrast, the distributions for $\tau_{\mathrm{highz}}$ are broad and consistent with zero within $2\sigma$, with extended tails reflecting residual degeneracy that allows for additional early ionization contributions. Across all values of $z_{\rm c}$, we find $\tau_{\mathrm{highz}} \lesssim 0.112$ (95\% CL), see Table~\ref{tab:tau_constraints}.

\textbf{\textit{Discussion and conclusions ---}} 
We apply our method to the case of a MeV-scale axion decaying around the epoch of reionization. The full phenomenological analysis is presented in the companion Paper~II~\cite{Yin:2025amn}. We consider a relic axion population of mass $m_a$ and photon coupling $g_{a\gamma\gamma}$, produced through Primakoff freeze-in at low reheating temperature~\cite{Cadamuro:2010cz,Cadamuro:2011fd}. The same coupling controls both axion production and the decay rate, allowing for a direct comparison between the energy injection history and our bounds on $\tau_{\mathrm{highz}}$.

Figure~\ref{fig:Xez} compares the standard $\Lambda$CDM ionization history with a model in which a subdominant axion component decays after recombination. For a minimal reheating temperature consistent with BBN, $T_{\rm reh} = 5$\,MeV~\cite{deSalas:2015glj, Hasegawa:2019jsa}, the axion constitutes a fractional DM abundance $\sim 10^{-7}$. The ionization fraction $X_e(z)$ is computed beyond the on-the-spot approximation, using the full energy deposition efficiency from \textsc{DarkHistory}~\cite{Liu:2019bbm}. For the benchmark parameters $m_a = 10^5$\,eV and $g_{a\gamma\gamma} = 6\times10^{-13}\,{\rm GeV^{-1}}$, the injected photons efficiently ionize the IGM at high redshift, producing an extreme optical depth $\tau_{\mathrm{reio}} \simeq 1.89$, far exceeding observational bounds. Figure~\ref{fig:tau(m,g)_chi} maps $\tau_{\mathrm{highz}}$ across the $(m_a, g_{a\gamma\gamma})$ plane. For $T_{\rm reh} = 5$\,MeV, the freeze-in process yields an effectively ``irreducible'' axion abundance~\cite{Langhoff:2022bij, Jain:2024dtw, Balazs:2022tjl}. The yellow contour shows the 95\% CL bound derived from our $\tau_{\mathrm{highz}}$ posterior, while the blue contour from Ref.~\cite{Langhoff:2022bij} represents CMB anisotropy limits derived using the on-the-spot approximation~\cite{Poulin:2016anj}. The red contour shows the low-redshift constraint from $\tau_{\rm lowz}$, consistent with the \textit{Planck} 2018 baseline~\cite{Planck:2018nkj}, and the white contour shows the exclusion obtained from a full $\chi^2$ analysis.

Our bound on $\tau_{\mathrm{highz}}$ is marginally more conservative than that from the on-the-spot approximation, but also more robust, as it employs the full energy deposition kernel. A direct comparison with the \textit{Planck} constraint on $\tau_{\mathrm{reio}}$, which implicitly assumes all contributions arise from low-redshift stellar reionization, would substantially underestimate the impact of early energy injection. By decomposing the total optical depth into $\tau_{\mathrm{lowz}}$ and $\tau_{\mathrm{highz}}$, our GP-based posterior provides a physically motivated and tighter constraint on high-$z$ ionization sources, particularly at large masses where early decays dominate.

\begin{figure}[htbp]
    \centering
    \includegraphics[width=1.1\linewidth]{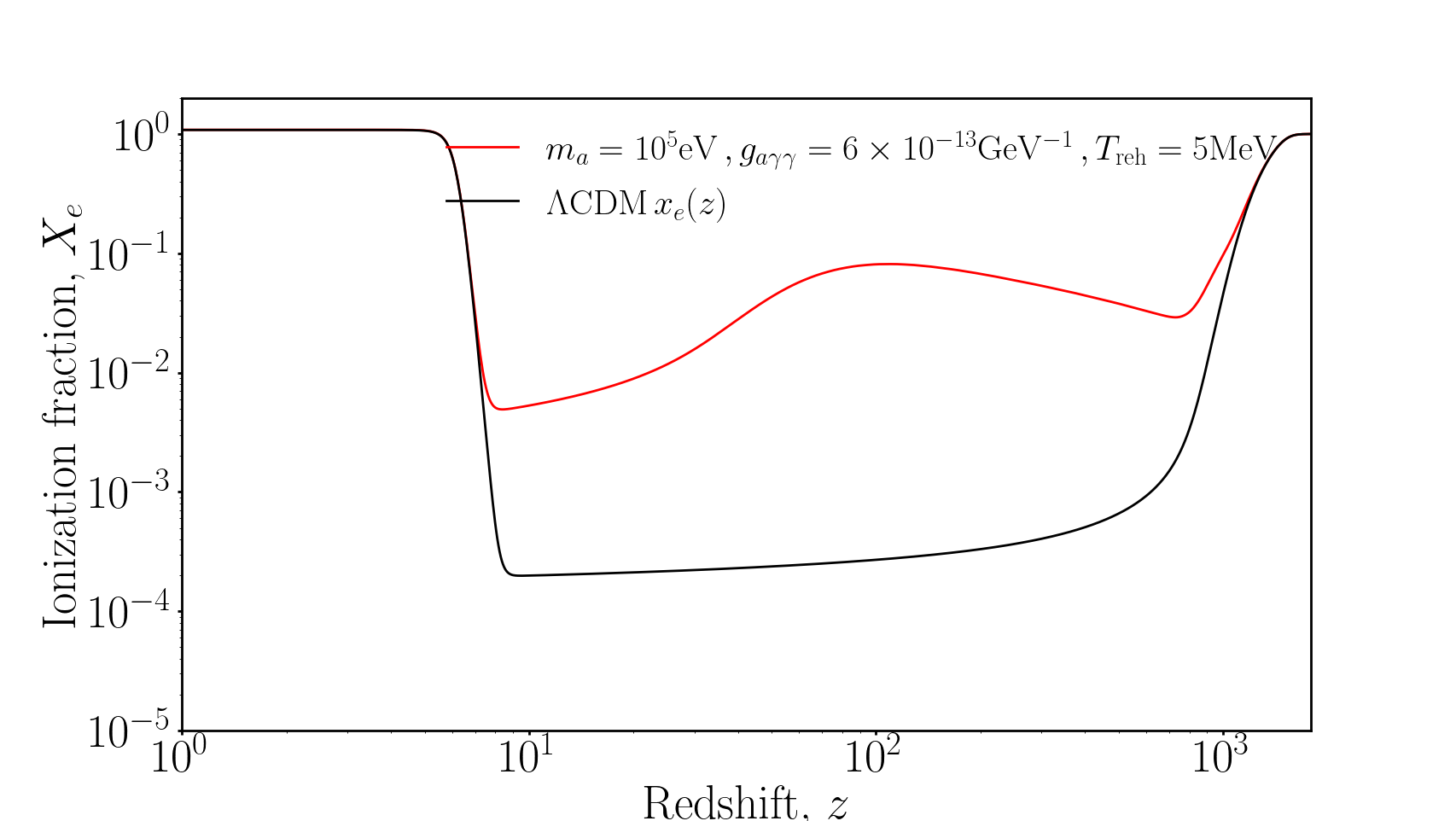}
\caption{Ionization fraction in the $\Lambda$CDM scenario compared to $\Lambda$CDM with one additional axion component, assuming instantaneous energy injection with the ``beyond on-the-spot'' assumption and using the full energy deposition efficiency function provided by \textsc{DarkHistory}~\cite{Liu:2019bbm}. In the model shown, the axion constitutes a fraction $\approx 10^{-7}$ of the DM (produced by freeze-in with the minimum reheating temperature consistent with BBN), before decaying at redshift $z_{\rm decay} < z_{\rm rec}$. The axion decay leads to efficient ionization of the IGM at high redshift, resulting in an unacceptably large CMB optical depth $\tau_{\mathrm{reio}} \approx 1.89$.}
    \label{fig:Xez}
\end{figure}

\begin{figure}[htbp]
    \centering
    \includegraphics[width=1\linewidth]{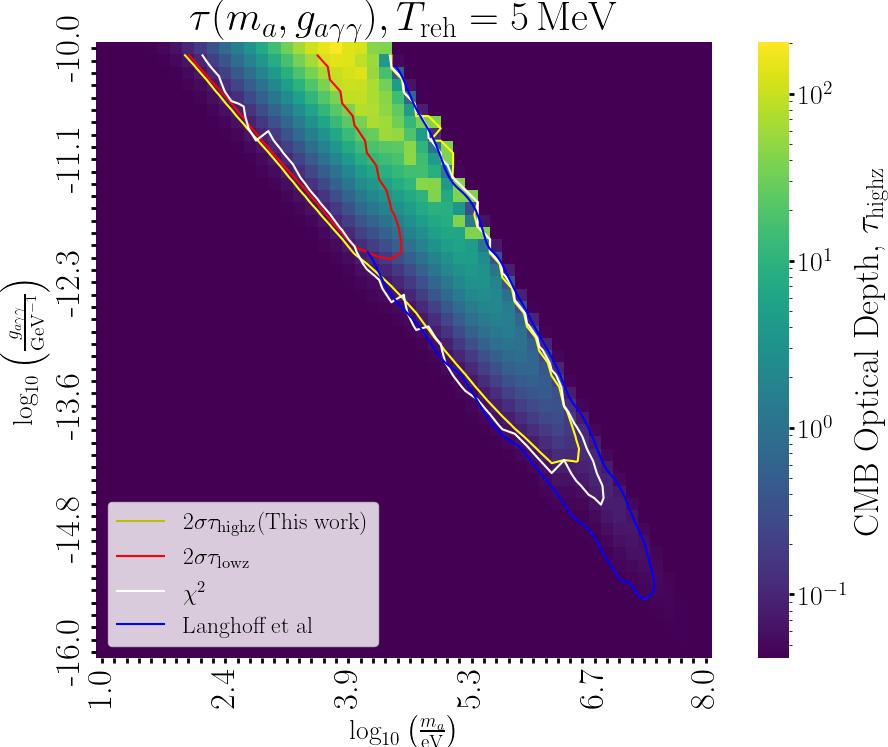}
    \caption{High-redshift optical depth $\tau_{\mathrm{highz}}$ (color scale) as a function of axion mass and axion–photon coupling. Solid lines show 95\%~CL exclusion contours from different methods: yellow from our GP-based $\tau_{\mathrm{highz}}$ reconstruction ($z \in [30,800]$), red from low-redshift constraints ($z \in [0,30]$) consistent with \textit{Planck}~2018~\cite{Planck:2018nkj}, blue from Ref.~\cite{Langhoff:2022bij} (``Langhoff et al.''), and white from the full $\chi^2$ analysis.}
    \label{fig:tau(m,g)_chi}
\end{figure}

We validate $\tau_{\mathrm{highz}}$ as a statistical proxy by comparing it to a full likelihood-based $\chi^2$ analysis. From our MCMC chains we determine the 95\% exclusion threshold and test 100 random decaying axion models from Paper~II. For each model, we compute $X_e(z)$, evaluate $\tau_{\mathrm{highz}}$, and calculate $\chi^2$ using the \emph{Planck} low-$\ell$ EE-only SimAll likelihood. A model is excluded if $\tau_{\mathrm{highz}}$ exceeds the upper limit in Table~\ref{tab:tau_constraints}, $\tau_{\mathrm{highz}} > 0.111$, or if it lies above the 95\% CL threshold of the $\chi^2$ distribution, $\chi^2 > 403.3$, leading to 21 models excluded by the $\tau_{\mathrm{highz}}$ criterion and 26 by the $\chi^2$ test. The $\chi^2$ exclusion contour (white in Fig.~\ref{fig:tau(m,g)_chi}) tracks the $\tau_{\mathrm{highz}}$ contour (yellow) at low masses, and becomes slightly tighter at high masses, confirming that $\tau_{\mathrm{highz}}$ offers a conservative yet reliable proxy for a full likelihood analysis.

The imprint of reionization on the CMB is a sensitive probe of exotic injection~\cite{Slatyer:2012yq, Liu:2019bbm, Poulin:2016anj, Langhoff:2022bij, Escudero:2023vgv}. Typically, such constraints are obtained model by model, requiring dedicated likelihood reanalyses. Our approach is complementary and model-independent. Using the GP-extended \texttt{CLASS} module, we generate non-parametric reionization histories $X_e(z)$ and constrain them with the \emph{Planck} low-$\ell$ EE-only SimAll likelihood up to $z=800$, extending previous GP-based reconstructions of the standard low-$z$ history. The implementation and posterior chains are publicly available at \href{https://github.com/Cheng-Hanyu/CLASS_reio_gpr}{github.com/Cheng-Hanyu/CLASS\_reio\_gpr}.

From this reconstruction, we define the derived parameter $\tau_{\mathrm{highz}}$, which isolates the high-redshift contribution to the CMB optical depth. This observable directly constrains exotic scenarios such as the ``irreducible'' freeze-in abundance of decaying MeV axions~\cite{Langhoff:2022bij}, giving bounds that closely reproduce those obtained from a full $\chi^2$ likelihood analysis. Being model-independent, the posterior on $\tau_{\mathrm{highz}}$ can be applied to any energy injection scenario for which $X_e(z)$ can be computed. In the companion Paper~II, we set limits on the reheating temperature in string theory compactifications. Using an ensemble of models constructed following Ref.~\cite{Gendler:2023kjt}, containing hundreds of axions in explicit Calabi--Yau realizations, we determine the maximal reheating temperature consistent with CMB reionization constraints across the landscape.

Our methods have several broader applications. The $\tau_{\mathrm{highz}}$ posterior can be used in artificial intelligence--driven searches of the string theory landscape, allowing for a quantitative link between reheating, geometry, and topology in observationally viable compactifications. Moreover, our results extend the analysis of Ref.~\cite{Escudero:2023vgv}, which sets constraints on ultralight axions heating the IGM through parametric resonance decay of merging axion stars into radio photons~\cite{Eby:2015hyx, Hertzberg:2020xdn, Du:2023jxh, Chung-Jukko:2023cow, Levkov:2020txo, Maseizik:2024qly, Di:2024tlz}. While that study used only the \emph{Planck} constraint on $\tau_{\rm lowz}$, our results show that doing so is overly conservative: limits derived from $\tau_{\mathrm{highz}}$ are substantially stronger.

More broadly, our analysis highlights the power of a model-independent approach to reionization. Current CMB polarization data already place strong constraints on high-redshift ionization and, by extension, on a wide range of exotic energy injection scenarios. By reconstructing the ionization history with Gaussian processes and introducing the derived parameter $\tau_{\mathrm{highz}}$, we provide a practical and robust tool to test early-Universe physics without relying on specific model templates. This framework transforms reionization into a precision probe of new physics, offering a clean and general way to assess the viability of scenarios beyond $\Lambda$CDM using existing data alone.

\begin{acknowledgments}
HC, ZY, and LV acknowledge support by the National Natural Science Foundation of China (NSFC) through the grant No.\ 12350610240 ``Astrophysical Axion Laboratories''. HC and ZY also acknowledge the support of the China Scholarship Council program (Project ID:202406230341 and 202406230357). EDV is supported by a Royal Society Dorothy Hodgkin Research Fellowship. DJEM is supported by an Ernest Rutherford Fellowship from the Science and Technologies Facilities Council, STFC (Grant No. ST/T004037/1), an STFC consolidator grant (Grant No.\ ST/X000753/1) and by a Leverhulme Trust Research Project (Grant No.\ RPG-2022-145). LV also thanks Istituto Nazionale di Fisica Nucleare (INFN) through the ``QGSKY'' Iniziativa Specifica project. LV additionally thanks the Tsung-Dao Lee Institute for hospitality during the final stages of this work. This publication is based upon work from the COST Actions ``COSMIC WISPers'' (CA21106) and ``Addressing observational tensions in cosmology with systematics and fundamental physics (CosmoVerse)'' (CA21136), both supported by COST (European Cooperation in Science and Technology). This work made use of the open source software matplotlib~\cite{2007CSE.....9...90H}, numpy~\cite{2020Natur.585..357H}, and scipy~\cite{2020NatMe..17..261V}.
\end{acknowledgments}

\bibliographystyle{apsrev4-1}
\bibliography{references.bib}

\clearpage
\newpage
\maketitle
\onecolumngrid
\begin{center}
\textbf{\large Constraining exotic high-\texorpdfstring{$z$}{z} reionization histories with Gaussian processes and the cosmic microwave background} \\ 
\vspace{0.1in}
{ \it \large Supplemental Material}\\ 
\vspace{0.05in}
{Hanyu Cheng, \ Ziwen Yin, \ Eleonora Di Valentino, \ David J. E.\ Marsh, and \ Luca Visinelli}
\end{center}
\onecolumngrid
\setcounter{equation}{0}
\setcounter{figure}{0}
\setcounter{table}{0}
\setcounter{section}{0}
\setcounter{subsection}{0}
\setcounter{page}{1}
\renewcommand{\theequation}{S\arabic{equation}}
\renewcommand{\thefigure}{S\arabic{figure}}
\renewcommand{\thetable}{S\arabic{table}}

\subsection{The reionization scheme \texttt{reio\_gpr\_tanh} in \texttt{CLASS}}
\label{appendix:parameters}

We summarize the method used to model the ionization fraction $X_e(z)$ at different redshifts, based on a trained Gaussian Process (GP). The procedure starts by specifying the input training data $X_e^{\mathrm{IN}}(z)$ and the corresponding redshifts. Using the trained GP, we generate an array of predictions $X_e^{\mathrm{GP}}(z)$ at selected interpolation points. A hyperbolic tangent (tanh) transition is then employed to smoothly connect neighboring GP-predicted $X_e^{\mathrm{GP}}(z)$ points. The \texttt{CLASS} code is modified by introducing a new reionization scheme, \texttt{reio\_gpr\_tanh}, controlled by the following parameters:
\begin{itemize}
    \item \textbf{\texttt{gpr\_reio\_num}}: Specifies the number of GP training points used to model the reionization history. Each training point corresponds to a redshift $z_i$, with a corresponding ionization fraction $X_e^{\mathrm{IN}}(z_i)$ at the $i$-th training point. These input pairs $\{z_i, X_e^{\mathrm{IN}}(z_i)\}$ form the dataset used to train the GP model for predicting $X_e(z)$ at arbitrary redshifts.

    \item \textbf{\texttt{gpr\_reio\_z}} and \textbf{\texttt{gpr\_reio\_xe}}: Arrays specifying the redshift values and corresponding ionization fraction values used for the training of the GP. These arrays together define the input training set $\{z_i,\, X_e^{\mathrm{IN}}(z_i)\}_{i=1}^{\texttt{gpr\_reio\_num}}$, which is used to build the ionization history $X_e(z)$ through GP regression.

    \item \textbf{\texttt{gpr\_z\_min}} and \textbf{\texttt{gpr\_z\_max}}: The minimum and maximum redshifts of the interpolation grid.

    \item \textbf{\texttt{gpr\_z\_transition}}: The redshift that separates the low- and high-redshift regions.

    \item \textbf{\texttt{gpr\_n\_low}}: The number of interpolation points in the low-redshift region, spanning from \textbf{\texttt{gpr\_z\_min}} to \textbf{\texttt{z\_transition}}.

    \item \textbf{\texttt{gpr\_n\_high}}: The number of interpolation points in the high-redshift region, spanning from \textbf{\texttt{z\_transition}} to \textbf{\texttt{gpr\_z\_max}}.

    \item \textbf{\texttt{gpr\_sigma\_f}} and \textbf{\texttt{gpr\_l}}: Hyperparameters of the Radial Basis Function (RBF) kernel used by the GP. The kernel is~\cite{JMLR:v11:chang10a}
    \begin{equation}
        k(z_{1}, z_{2}) \;=\; \sigma_\mathrm{f}^{2} \exp \Bigl[
        -\tfrac{(z_{1} - z_{2})^{2}}{2\,l^{2}}
        \Bigr],
    \label{eq:rbf-kernel}
    \end{equation}
    where $\sigma_{\mathrm{f}} = \texttt{gpr\_sigma\_f}$ sets the overall variance scale, and $l = \texttt{gpr\_l}$ is the correlation length in redshift space.

    \item \textbf{\texttt{gpr\_reio\_step\_sharpness}}: Controls the sharpness of the tanh transitions used to smoothly interpolate between neighboring GP-predicted ionization fraction values. Smaller values correspond to sharper, more step-like transitions, while larger values yield smoother, more gradual changes in $X_e^{\mathrm{GP}}(z)$.
\end{itemize}

\subsection{Gaussian process model}
\label{appendix:GP}

The GP model, denoted \texttt{gp\_model}, is the foundation of our reionization scheme. After reading the training points $\bigl\{z_i,\, X_e^{\mathrm{IN}}(z_i)\bigr\}$ from the input, a mean function $\bar{m}(z)$ can be specified for the GP prior. In this paper, we adopt a zero-mean prior:
\begin{equation}
    \bar{m}(z) \;=\; 0\,.
\label{eq:mean-zero}
\end{equation}
For the kernel, we use the standard RBF kernel shown in Eq.~\eqref{eq:rbf-kernel}. Given $N$ training points, we construct the $N \times N$ kernel matrix $\mathbf{K}$ with entries
\begin{equation}
    \mathbf{K}_{ij} = k(z_i, z_j)\,,
\end{equation}
where $k$ is the RBF kernel evaluated at redshifts $z_i$ and $z_j$. For any redshift $z$, the ionization fraction is predicted as
\begin{equation}
    X_e^{\mathrm{GP}}(z)
    \;=\;
    \bar{m}(z)
    \;+\;
    \mathbf{k}_{*}^{T}\,
    \mathbf{K}^{-1}\,
    \Bigl[
        \mathbf{X}_{\mathrm{train}}
        \;-\;
        \bar{m}\bigl(\mathbf{z}_{\mathrm{train}}\bigr)
    \Bigr],
\label{eq:gp-mean}
\end{equation}
where $\mathbf{k}_\mathrm{*}$ is the vector of covariances between a new point $z$ and each training redshift $z_i$, so that $\mathbf{k}_\mathrm{*,\,i} = k(z, z_i)$. The matrix $\mathbf{K}^{-1}$ is the inverse of the kernel matrix $\mathbf{K}$. We define the training data array as $\mathbf{X}_{\mathrm{train}} \equiv \bigl[X_e^{\mathrm{IN}}(z_{1}), X_e^{\mathrm{IN}}(z_{2}), \ldots, X_e^{\mathrm{IN}}(z_\mathrm{N})\bigr]^{T}$, while $\bar{m}(\mathbf{z}_{\mathrm{train}})$ is the mean function evaluated at the training points. Here, the mean function is set to zero.
Once the GP is trained, we generate \textbf{\texttt{n\_gpr}} interpolation grid points, see Eq.~\eqref{eq:total_points}, ranging from \texttt{gpr\_z\_min} to \texttt{gpr\_z\_max}, using the adaptive binning strategy. For each grid point $z_{\mathrm{grid}}$, the ionization fraction $X_e^{\mathrm{GP}}(z_{\mathrm{grid}})$ is predicted according to Eq.~\eqref{eq:gp-mean}, with a lower bound imposed to ensure physically meaningful, non-negative values. Here, we set the lowest value of the ionization fraction to $10^{-5}$, as in Eq.~\eqref{eq:boundary-upper} below.

We discuss our choice for the hyperbolic tangent transition. After computing the GP predictions over the interpolation grid, we discretize the reionization history into bins, following the approach used in \texttt{reio\_bins\_tanh} as defined in the \texttt{CLASS} documentation. For each pair of consecutive redshifts $z_i$ and $z_\mathrm{i+1}$, with corresponding GP-predicted ionization fractions $X_e^{\mathrm{GP}}(z_i)$ and $X_e^{\mathrm{GP}}(z_\mathrm{i+1})$, the ionization fraction is smoothly interpolated using a tanh transition. Specifically, we set
\begin{equation}
\begin{split}
    X_e(z)
    =
    X_e^{\mathrm{GP}}(z_{i}) +
    \tfrac{1}{2}\,\Bigl[
        1 + \tanh\!\Bigl(\tfrac{z - z_{\mathrm{jump}}}{\Delta z}\Bigr)
    \Bigr]
    \Bigl[
        X_e^{\mathrm{GP}}(z_{i+1}) \!-\! X_e^{\mathrm{GP}}(z_{i})
    \Bigr],
\label{eq:tanh-transition}
\end{split}
\end{equation}
where $z_{\mathrm{jump}}$ is a suitably chosen midpoint between $z_i$ and $z_\mathrm{i+1}$, and $\Delta z = \texttt{gpr\_reio\_step\_sharpness}$ controls the sharpness of the transition.

We now discuss boundary conditions, which are imposed at the lower and upper ends of the redshift range. For small $z$, corresponding to the post-reionization epoch, the code enforces a fully ionized Universe, with
\begin{equation}
    X_e \;\approx\; 1 \;+\;
    \frac{Y_{\mathrm{He}}}{4\,(1 - Y_{\mathrm{He}})}
    \,,
\label{eq:boundary-lower}
\end{equation}
to account for ionized hydrogen and singly ionized helium, with mass fraction $Y_{\mathrm{He}}$. This expression can be further modified to include double ionization of helium, if relevant, as in Eq.~\eqref{eq:full_ionization}. At high redshifts, corresponding to the pre-reionization epoch, the ionization fraction is set to a small floor value to ensure that the Universe is essentially neutral at very early times. Here, we set
\begin{equation}
    X_e \;\approx\; 10^{-5}\,.
\label{eq:boundary-upper}
\end{equation}
This value ensures that the model remains physical across the entire redshift range.

Additionally, this approach aims to balance the thorough exploration of the entire parameter space of $X_e(z)$ and $z$ with computational efficiency during MCMC analyses. By implementing this adaptive binning strategy, we reduce the computational burden without significantly compromising reconstruction accuracy.

The total number of interior interpolation points is computed as:
\begin{equation}
    \textbf{\texttt{n\_gpr}} = \textbf{\texttt{gpr\_n\_low}} + \textbf{\texttt{gpr\_n\_high}} - 1\,,
    \label{eq:total_points}
\end{equation}
where the subtraction of 1 ensures that the transition point is not double-counted. The bin sizes in the two regions are calculated by:
\begin{equation}
\label{eq:dz_lowhigh}
\begin{split}
    \textbf{\texttt{dz\_low}} &= \frac{\textbf{\texttt{z\_transition}} - \textbf{\texttt{gpr\_z\_min}}}{\textbf{\texttt{gpr\_n\_low}} - 1}\,,\\
    \textbf{\texttt{dz\_high}} &= \frac{\textbf{\texttt{gpr\_z\_max}} - \textbf{\texttt{z\_transition}}}{\textbf{\texttt{gpr\_n\_high}} - 1}\,.
\end{split}
\end{equation}

Besides the priors listed in Table~\ref{tab:priors}, we input 20 redshift values ($z_1$ to $z_{20}$) and 20 ionization fraction values ($X_{e,1}$ to $X_{e,20}$). The prior ranges are set as follows:
\begin{itemize}
    \item For $z_1$, the prior range is $[4, 8]$.
    \item For $z_2$, the prior range is $[8.1, 25]$.
    \item For $z_3$ to $z_{20}$, the values evenly cover the range from $25.1$ to nearly $800$, ensuring that the bins do not overlap.
    \item For $X_{e,1}$ to $X_{e,20}$, the prior range is $[10^{-5}, 1.05]$. Specifically, $X_{e,1}$ is centered at $1.05$, $X_{e,2}$ is centered at $0.1$, and for $X_{e,3}$ to $X_{e,20}$ we first define $\log_{10}(X_e)$ with the center value set to $\log_{10}(X_e) = -5$. The values of the reionization fraction are then passed to \textbf{\texttt{gpr\_reio\_xe}}.
\end{itemize}

\subsection{Validating the standard reionization against \textit{Planck} 2018 data}
\label{appendix:validation}

\begin{figure*}[htp]
\setkeys{Gin}{width=0.48\linewidth}
    \subfloat{\includegraphics[height=7cm]{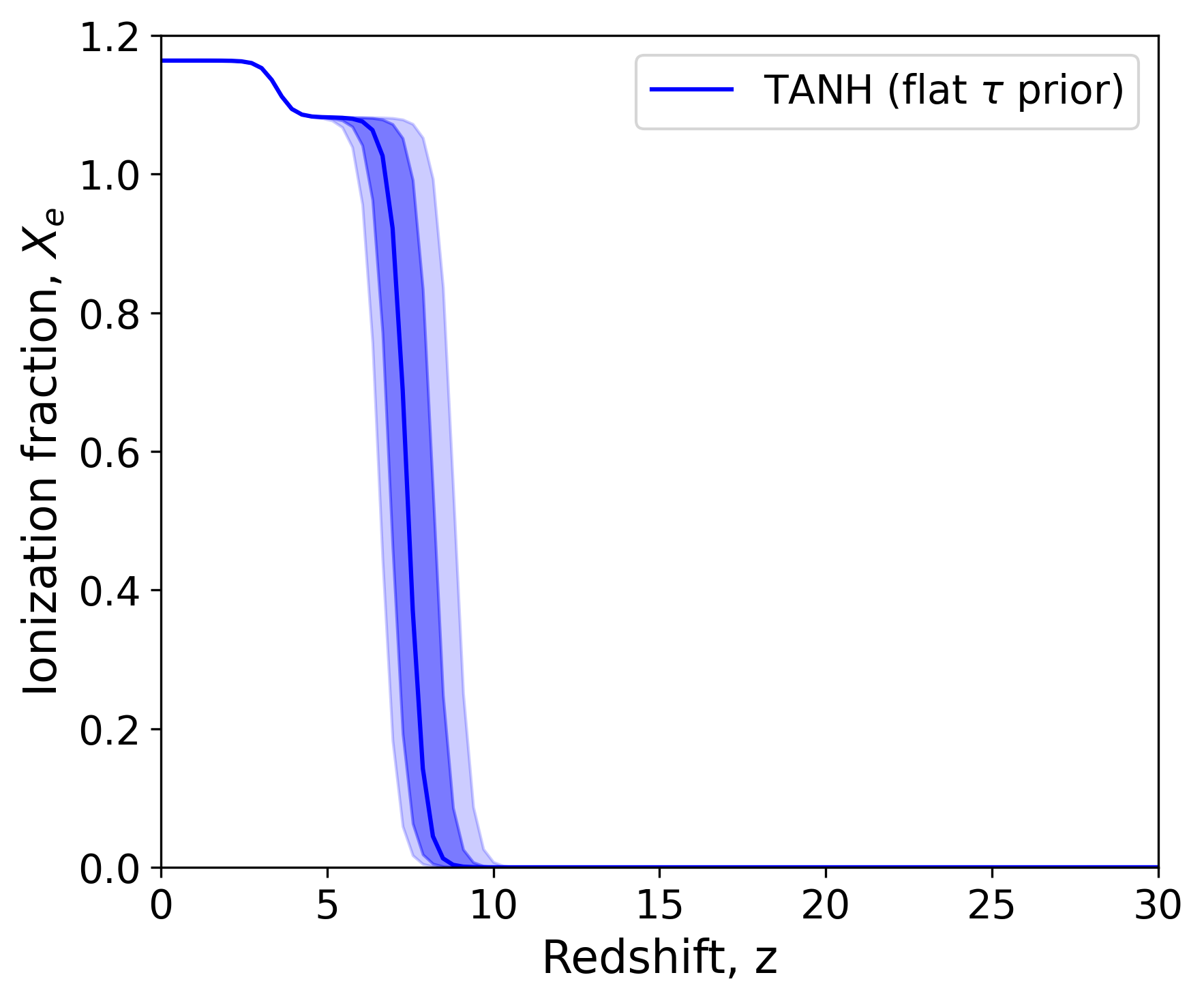}}\hfill
    \subfloat{\includegraphics[height=7cm]{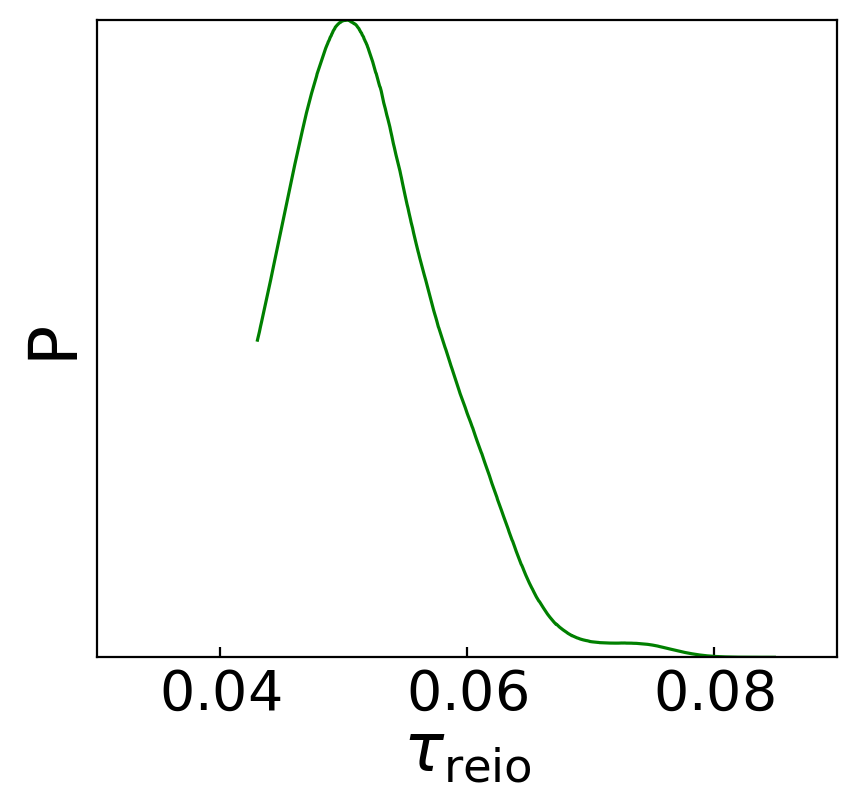}}
    \caption{The free electron fraction $X_e(z)$ (left) and marginalized constraints on the optical depth $\tau_{\mathrm{reio}}$ (right), obtained using lowE data alone. The amplitude of scalar perturbations, $A_{\mathrm{s}}\,e^{-2 \tau_{\mathrm{reio}}}$, as well as other cosmological parameters, are fixed to their best-fit values from the \textit{Planck} TT,TE,EE+lowE datasets. A flat prior on $\tau_{\mathrm{reio}}$ is adopted, with a lower bound of $\tau_{\mathrm{reio}} > 0.043$ imposed to enforce observational constraints from the Gunn-Peterson effect.}
    \label{fig:constraint_tau_Xe}
\end{figure*}

Figure~\ref{fig:constraint_tau_Xe} presents marginalized constraints on the optical depth, $\tau_{\mathrm{reio}}$, and the free electron fraction, $X_e(z)$, derived using the lowE dataset alone, with all other cosmological parameters fixed to their best-fit values from the \textit{Planck} 2018 TT,TE,EE+lowE analysis~\cite{Planck:2018vyg}. A flat prior on $\tau_{\mathrm{reio}}$ is adopted, with a lower bound of $\tau_{\mathrm{reio}} > 0.043$ imposed to enforce observational constraints from the Gunn-Peterson effect. This lower limit ensures consistency with observations of high-redshift quasar spectra, which demonstrate that the IGM is highly ionized by $z \approx 6$~\cite{Fan:2006dp, Bouwens:2015vha}. The imposed bound effectively requires reionization to be substantially complete by this redshift, preventing unphysically late reionization scenarios that would conflict with these well-established observations. The results illustrate the sensitivity of the low-multipole polarization data to the integrated reionization history, while emphasizing the limited ability of current data to resolve detailed features in $X_e(z)$ without additional assumptions or external information.

\subsection{Equally spaced interpolation bins}
\label{appendix:equally_spaced_bins}

Here, we introduce an alternative method that uses equally spaced redshift points for interpolation. Unlike the approach described in Sec.~IIA, this method does not require the parameters \textbf{\texttt{gpr\_z\_transition}}, \textbf{\texttt{gpr\_n\_low}}, and \textbf{\texttt{gpr\_n\_high}}. Instead, it uses only a single parameter, \textbf{\texttt{gpr\_n\_interp}}, which specifies the number of equally spaced interpolation points, while all other parameters remain the same as in Sec.~IIA. This method is more suitable for scenarios involving a relatively narrow redshift range—for example, when the maximum redshift is below $z \lesssim 50$. We have performed a variety of tests using this approach to evaluate its reliability.

\begin{itemize}
    \item 
    \textbf{\texttt{gpr\_n\_interp}}: Specifies the number of equally spaced redshift points at which the trained GP model is used to predict the ionization fraction $X_e^{\mathrm{GP}}(z)$. The code constructs a uniform interpolation grid $\{z_{\mathrm{grid}}\} \in [\texttt{gpr\_z\_min},\, \texttt{gpr\_z\_max}]$ consisting of \texttt{gpr\_n\_interp} points, and evaluates the trained GP model at each redshift in this grid. These predicted values serve as interpolation nodes for reconstructing the continuous ionization history.
\end{itemize}

\begin{table}[htbp]
\centering
\begin{tabular}{lcc}
\toprule
Case & $\tau_{\mathrm{reio}}$ & $R-1$ \\
\midrule
\texttt{recoverplanck}            & $0.0531^{+0.0027}_{-0.0099}$ & $0.0009$ \\
\texttt{reio\_camb}               & $0.0497^{+0.0082}_{-0.0073}$ & $0.01$   \\
\texttt{bins\_two\_z\_z\_free}    & $0.0497^{+0.0038}_{-0.0053}$ & $0.006$  \\
\texttt{bins\_two\_z\_z\_fixed}   & $0.0506^{+0.0026}_{-0.0030}$ & $0.008$  \\
\texttt{gpr\_two\_z\_z\_fixed}    & $0.0506\pm 0.0025$           & $0.008$  \\
\texttt{gpr\_fixtwoz}             & $0.0542\pm 0.0026$           & $0.02$   \\
\texttt{gpr\_fixfourz}            & $0.0544\pm 0.0027$           & $0.07$   \\
\texttt{gpr\_fixhyper}            & $0.0537\pm 0.0045$           & $0.01$   \\
\texttt{gpr\_free}                & $0.0525^{+0.0057}_{-0.0041}$ & $0.07$   \\
\texttt{gpr\_largerange}          & $0.0511\pm 0.0054$           & $0.06$   \\
\texttt{gpr\_larger}              & $0.0521^{+0.0051}_{-0.0057}$ & $0.03$   \\
\texttt{gpr\_doubleYHe}           & $0.0549\pm 0.0048$           & $0.04$   \\
\texttt{gpr\_fixonexe\_doubleYHe} & $0.0533\pm 0.0057$           & $0.02$   \\
\bottomrule
\end{tabular}
\caption{Results of data analysis using the \textit{Planck} 2018 low-$\ell$ EE likelihood. The first column reports different cases with various reionization parameterizations and prior choices. The second column lists the best-fit value of $\tau_{\mathrm{reio}}$ with \textit{Planck} constraints at 68\% CL. The third column presents the convergence factor $R - 1$, where smaller values indicate better convergence. All cases show satisfactory convergence.}
\label{tab:results}
\end{table}

\begin{figure}[htbp]
    \centering
    \includegraphics[width=0.7\linewidth]{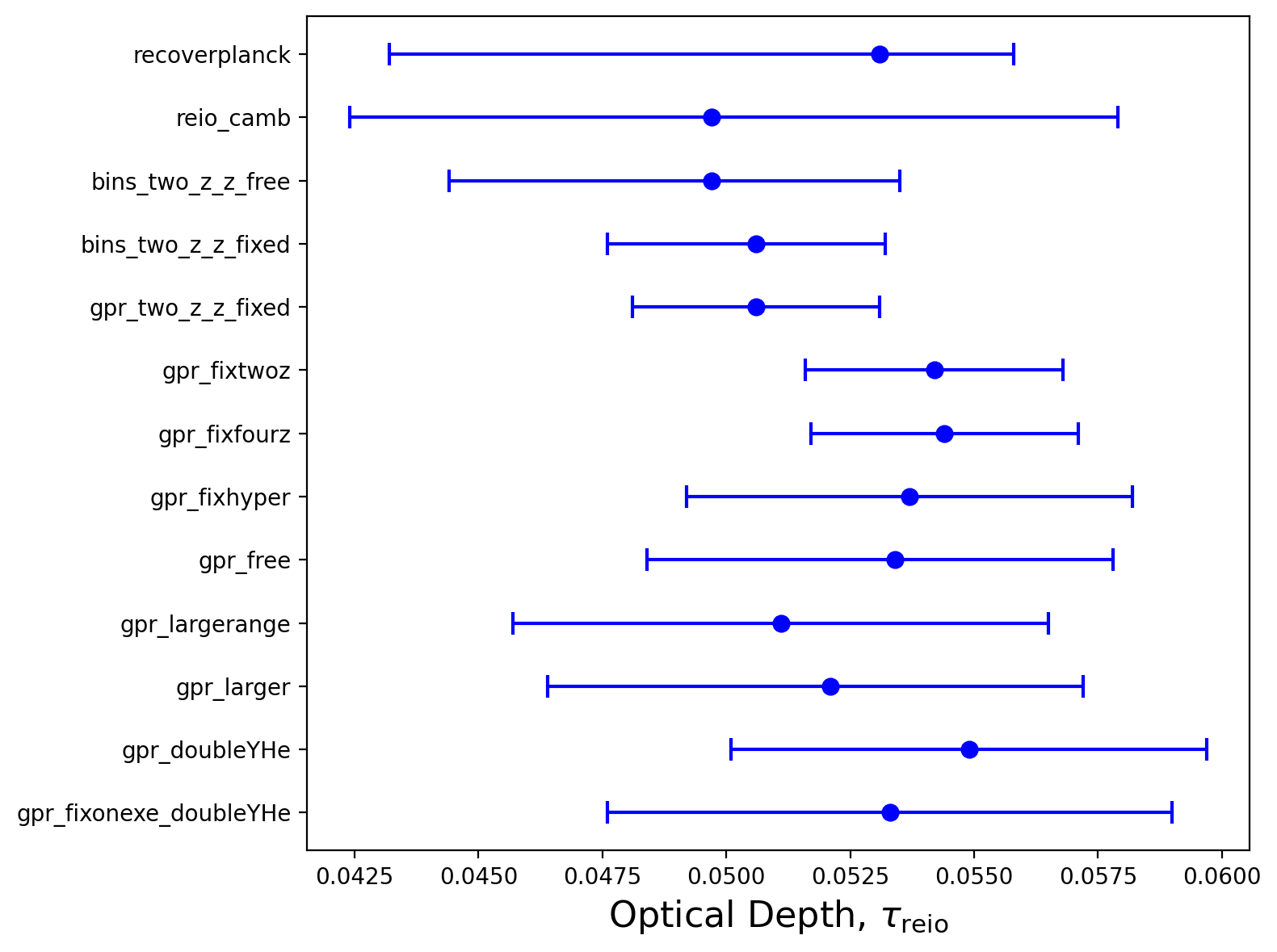}
    \caption{Whisker plot displaying the optical depth $\tau_{\mathrm{reio}}$ values from different reionization parameterizations and prior choices, visualizing the results presented in Table~\ref{tab:results}.}
    \label{fig:whisker_plot}
\end{figure}

The results summarized in Table~\ref{tab:results} indicate that the best-fit values of the optical depth $\tau_{\mathrm{reio}}$ are largely consistent across the different cases, as also illustrated in the whisker plot in Fig.~\ref{fig:whisker_plot}. Notably, the identical values of $\tau_{\mathrm{reio}}$ found for the \texttt{bins\_two\_z\_z\_fixed} and \texttt{gpr\_two\_z\_z\_fixed} cases demonstrate that the newly introduced \texttt{reio\_gpr\_tanh} parameterization is in excellent agreement with the previously used \texttt{reio\_bins\_tanh} parameterization.

\begin{table}[htbp]
    \centering
    \begin{tabular}{llp{8cm}}
    \toprule
    Case & Reionization Case & Prior \\
    \midrule
    \texttt{recoverplanck} &
    reio\_camb &
    \parbox[t]{\linewidth}{
    $\tau_{\mathrm{reio}}:[0.043,\,0.8]$
    } \\[2mm]

    \texttt{reio\_camb} &
    reio\_camb &
    \parbox[t]{\linewidth}{
    $\tau_{\mathrm{reio}}:[0.01,\,0.8]$
    } \\[2mm]

    \texttt{bins\_two\_z\_z\_free} &
    reio\_bins\_tanh &
    \parbox[t]{\linewidth}{
    $z_{1}:[4.5,\,5.5]$\\
    $z_{2}:[8,\,11]$\\
    } \\[2mm]

    \texttt{bins\_two\_z\_z\_fixed} &
    reio\_bins\_tanh &
    \parbox[t]{\linewidth}{
    ${\rm binned\_reio}\_z:(5.0,\,9.5)$\\
    } \\[2mm]

    \texttt{gpr\_two\_z\_z\_fixed} &
    reio\_gpr\_tanh &
    \parbox[t]{\linewidth}{
    ${\rm gpr\_reio\_z}:(5.0,\,9.5)$\\
    ${\rm gpr\_z_{\mathrm{min}}}:5.0$\\
    ${\rm gpr\_z_{\mathrm{max}}}:9.5$\\
    } \\[2mm]

    \texttt{gpr\_fixtwoz} &
    reio\_gpr\_tanh &
    \parbox[t]{\linewidth}{
    ${gpr\_reio\_z}:(5.5,\,10.0)$\\
    $X_{e1}:[0.95,\,1.05]$\\
    $X_{e2}:[0.0001,\,0.2]$\\
    } \\[2mm]

    \texttt{gpr\_fixfourz} &
    reio\_gpr\_tanh &
    \parbox[t]{\linewidth}{
    ${\rm gpr\_reio\_num}:4$\\
    ${\rm gpr\_n\_interp}:2$\\
    ${\rm gpr\_reio\_z}:(5.5,\,7.0,\,8.5,\,10.0)$\\
    $X_{e1}:[0.95,\,1.05]$\\
    $X_{e2}:[0.55,\,0.9]$\\
    $X_{e3}:[0.25,\,0.5]$\\
    $X_{e4}:[0.0001,\,0.2]$\\
    } \\[2mm]

    \texttt{gpr\_fixhyper} &
    reio\_gpr\_tanh &
    \parbox[t]{\linewidth}{
    ${\rm gpr\_reio\_step\_sharpness}:0.3$\\
    ${\rm gpr\_sigma\_f}:0.3$\\
    ${\rm gpr\_l}:2.0$\\
    } \\[2mm]

    \texttt{gpr\_free} &
    reio\_gpr\_tanh &
    \parbox[t]{\linewidth}{
    ${\rm gpr\_reio\_step\_sharpness}:[0.2,\,0.4]$\\
    ${\rm gpr\_sigma\_f}:[0.2,\,0.4]$\\
    ${\rm gpr\_l}:[1.5,\,2.5]$\\
    } \\[2mm]

    \texttt{gpr\_largerange} &
    reio\_gpr\_tanh &
    \parbox[t]{\linewidth}{
    ${\rm gpr\_reio\_step\_sharpness}:[0.1,\,0.5]$\\
    ${\rm gpr\_sigma\_f}:[0.1,\,0.5]$\\
    ${\rm gpr\_l}:[1.0,\,3.0]$\\
    } \\[2mm]

    \texttt{gpr\_larger} &
    reio\_gpr\_tanh &
    \parbox[t]{\linewidth}{
    ${\rm gpr\_reio\_step\_sharpness}:[0.1,\,0.7]$\\
    ${\rm gpr\_sigma\_f}:[0.1,\,0.7]$\\
    ${\rm gpr\_l}:[1.0,\,5.0]$\\
    } \\[2mm]
    
    \texttt{gpr\_doubleYHe} &
    reio\_gpr\_tanh &
    \parbox[t]{\linewidth}{
    ${\rm gpr\_reio\_step\_sharpness}:[0.2,\,0.4]$\\
    ${\rm gpr\_sigma\_f}:[0.2,\,0.4]$\\
    ${\rm gpr\_l}:[1.5,\,2.5]$\\
    } \\[2mm]
    
    \texttt{gpr\_fixonexe\_doubleYHe} &
    reio\_gpr\_tanh &
    \parbox[t]{\linewidth}{
    $X_{e1}:[1.0,\,1.1]$\\
    $X_{e2}:0.0001$
    } \\
    \bottomrule
    \end{tabular}
    \caption{Detailed description of the reionization cases and corresponding prior ranges. The first column lists the case name, the second column gives the reionization parameterization used, and the third column provides the prior ranges for some parameters. Other common prior range choices are listed in the text. Here, parentheses represent the input training array for redshift $z$, while square brackets represent the prior range for different parameters.}
    \label{tab:priors1}
\end{table}

Additionally, comparing the \texttt{gpr\_fixtwoz} and \texttt{gpr\_fixfourz} cases reveals that the estimated value of $\tau_{\mathrm{reio}}$ is not significantly affected by the number of input redshift training points. Slight variations in $\tau_{\mathrm{reio}}$ among other cases reflect the influence of differing reionization parameterizations, whether redshift points are fixed or free, and the specific prior range chosen. For instance, cases with fixed redshift points, such as \texttt{gpr\_fixtwoz} or \texttt{gpr\_fixfourz}, tend to yield marginally larger values of $\tau_{\mathrm{reio}}$ and tighter error bars, while freeing all parameters (such as in \texttt{gpr\_free}, \texttt{gpr\_largerange}, and \texttt{gpr\_larger}) tends to result in larger error bars. The imposition of a fully ionized condition—as in the \texttt{gpr\_doubleYHe} and \texttt{gpr\_fixonexe\_doubleYHe} cases—leads to an increase in $\tau_{\mathrm{reio}}$, owing to the higher post-reionization electron fraction.

For the fully ionized condition adopted in the \texttt{gpr\_doubleYHe} and \texttt{gpr\_fixonexe\_doubleYHe} cases, we use the relation
\begin{equation}
X_e \approx 1 + 2\,\frac{Y_{\mathrm{He}}}{4\,(1 - Y_{\mathrm{He}})},
\label{eq:full_ionization}
\end{equation}
which accounts for the additional free electrons contributed by helium reionization, in contrast with Eq.~\eqref{eq:boundary-lower}, used for the analysis in the main body of the paper.

In addition to the detailed prior ranges provided in Table~\ref{tab:priors1}, we adopt several specific choices in our analysis. With the exception of the \texttt{gpr\_fixfourz} case, the parameters \texttt{gpr\_reio\_num} and \texttt{gpr\_n\_interp} are fixed to 2 for all scenarios employing the \texttt{reio\_gpr\_tanh} reionization model. For the \texttt{recoverplanck} case, we impose a lower bound on the optical depth as $\tau_{\mathrm{reio}} > 0.043$, following the \textit{Planck} 2018 treatment. In contrast, the \texttt{reio\_camb} case adopts a flat prior of $\tau_{\mathrm{reio}} \in [0.01,\,0.8]$ with no additional constraints. For the cases \texttt{bins\_two\_z\_z\_free}, \texttt{bins\_two\_z\_z\_fixed}, and \texttt{gpr\_two\_z\_z\_fixed}, identical priors are chosen for the electron fraction parameters, namely
\begin{equation}
    X_{e1}:[0.95,\,1.05]\,, \quad \text{and} \quad X_{e2}:[0.01,\,0.2]\,.
\end{equation}
In the \texttt{bins\_two\_z\_z\_free} case, both the redshift $z$ and the electron fraction $X_e$ vary freely. In contrast, for the \texttt{bins\_two\_z\_z\_fixed} and \texttt{gpr\_two\_z\_z\_fixed} cases, we fix the redshift $z$ while allowing $X_e$ to vary. This design enables us to test whether the new \texttt{reio\_gpr\_tanh} parameterization is consistent with the \texttt{reio\_bins\_tanh} model employed in the \texttt{CLASS} documentation. 

For the \texttt{bins\_two\_z\_z\_fixed}, \texttt{gpr\_two\_z\_z\_fixed}, and \texttt{gpr\_fixhyper} cases, the GP hyperparameters are fixed at
\begin{equation}
    {\rm gpr\_reio\_step\_sharpness = 0.3\,,\quad gpr\_sigma\_f = 0.3\,,\quad gpr\_l = 2.0}\,.
\end{equation}
For the \texttt{gpr\_fixtwoz} and \texttt{gpr\_fixfourz} cases, the following priors are used:
\begin{equation}
\begin{split}
    &{\rm gpr\_reio\_step\_sharpness:[0.2,\,0.4], \quad gpr\_sigma\_f:[0.2,\,0.4], \quad gpr\_l:[1.5,\,2.5]}\,,\\
    &{\rm gpr\_z\_min:[5.0,\,7.0], \quad gpr\_z\_max:[8.5,\,10.5]}\,.
\end{split}
\end{equation}
These two cases differ in the number of fixed redshift points, namely two in \texttt{gpr\_fixtwoz} and four in \texttt{gpr\_fixfourz}, to examine whether the number of fixed redshift points influences the estimated value of $\tau_{\mathrm{reio}}$.

For the \texttt{gpr\_free}, \texttt{gpr\_largerange}, and \texttt{gpr\_larger} cases, all parameters are allowed to vary, with the prior ranges enlarged in the latter two cases. For the \texttt{gpr\_fixhyper}, \texttt{gpr\_free}, and \texttt{gpr\_doubleYHe} cases, we adopt the same priors for the redshift and electron fraction parameters:
\begin{equation}
\begin{split}
    &{\rm gpr\_z\_min:[5.0,\,7.0], \quad gpr\_z\_max:[8.5,\,10.5]}\,,\\
    &z_1:[5.0,\,7.5]\,, \qquad z_2:[8.5,\,11.5]\,,\\
    &X_{e1}:[0.95,\,1.05], \quad X_{e2}:[0.0001,\,0.2]\,.
\end{split}
\end{equation}
For the \texttt{gpr\_largerange} and \texttt{gpr\_larger} cases, the priors are chosen as follows:
\begin{equation}
\begin{split}
    &{\rm gpr\_z\_min:[4.5,\,7.5], \quad gpr\_z\_max:[8.0,\,11.0]}\,,\\
    &z_1:[4.5,\,7.5]\,, \qquad z_2:[8.0,\,11.5]\,,\\
    &X_{e1}:[0.95,\,1.05], \quad X_{e2}:[0.0001,\,0.2]\,.
\end{split}
\end{equation}
For the \texttt{gpr\_fixonexe\_doubleYHe} case, we adopt the same priors as in the \texttt{gpr\_larger} case, except for the electron fraction parameters $X_{e1}$ and $X_{e2}$, which follow different prior ranges. Moreover, in this case we fix the electron fraction to $X_e = 10^{-4}$ at the higher redshift point $z_{2}$, enforcing a single tanh transition in the ionization history.

\subsection{The total optical depth}
\label{appendix:tautotal}

\begin{figure}[htbp]
    \centering
    \includegraphics[width=0.5\linewidth]{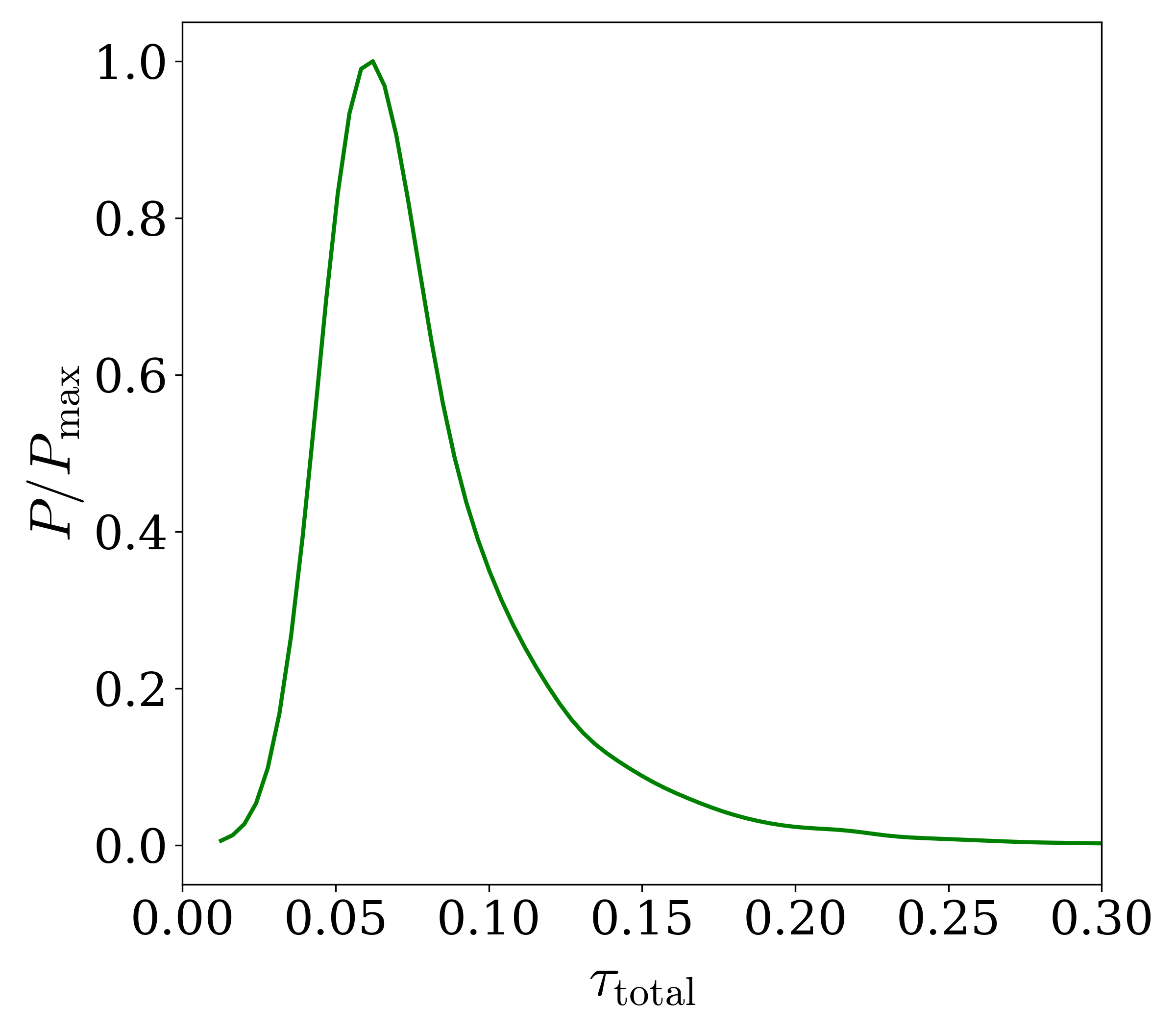}
    \caption{Normalized posterior distribution, $P/P_{\mathrm{max}}$, for the total optical depth $\tau_{\mathrm{reio}} = \tau_{\mathrm{lowz}} + \tau_{\mathrm{highz}}$. We find $\tau_{\mathrm{reio}} = 0.071^{+0.043}_{-0.019}$ at 68\% CL and $\tau_{\mathrm{reio}} = 0.071^{+0.119}_{-0.031}$ at 95\% CL. Results are derived from \textit{Planck} low-$\ell$ EE data alone.}
    \label{fig:tau_total}
\end{figure}

We report the total reionization optical depth, defined as in Eq.~\eqref{eq:tau_reio}. Our analysis yields $\tau_{\mathrm{total}} = 0.070^{+0.041}_{-0.018}$ (68\% CL) and $\tau_{\mathrm{total}} = 0.070^{+0.12}_{-0.030}$ (95\% CL), assuming an upper redshift limit of $z_{\mathrm{max}} = 800$. The normalized posterior distribution for $\tau_{\mathrm{total}}$ is shown in Fig.~\ref{fig:tau_total}.

For a direct comparison with constraints derived from high-$\ell$ \textit{Planck} data~\cite{Giare:2023ejv}, we also compute the optical depth using the standard definition of $\tau_{\mathrm{reio}}$ as implemented in the \texttt{CLASS} Boltzmann code. In this convention, the optical depth is integrated from $z = 0$ up to the redshift corresponding to the global minimum of the free electron fraction $X_e(z)$, marking the onset of reionization. Adopting this definition, we find
\begin{equation}
    \begin{split}
    \tau_{\mathrm{reio}} = 0.064^{+0.004}_{-0.027} \quad \hbox{(68\% CL)}\,,\\
    \tau_{\mathrm{reio}} = 0.064^{+0.063}_{-0.038} \quad \hbox{(95\% CL)}\,,
    \end{split}
\end{equation}
with the corresponding posterior distribution shown in Fig.~\ref{fig:tau_reio}. This result agrees well with the recent determination from high-$\ell$ \textit{Planck} data, $\tau_{\mathrm{reio}} = 0.080 \pm 0.012$ (68\%~CL)~\cite{Giare:2023ejv}.

A slightly higher value of $\tau_{\mathrm{reio}}$ can help to alleviate several well-known CMB anomalies, such as:  
(i)~the excess of lensing power quantified by $A_{\mathrm{lens}}$~\cite{Planck:2018vyg, DiValentino:2019qzk, Addison:2023fqc};  
(ii)~the preference for a closed Universe ($\Omega_k < 0$)~\cite{Planck:2018vyg, DiValentino:2019qzk, Handley:2019tkm};  
(iii)~the tight upper bound on the total neutrino mass, which is in tension with oscillation data~\cite{Jiang:2024viw, Jhaveri:2025neg};  
(iv)~the mild preference for an evolving dark energy component~\cite{Sailer:2025lxj}; and  
(v)~the tension between \textit{Planck} and JWST measurements within the $\Lambda$CDM framework~\cite{Forconi:2023izg}.
More generally, a higher optical depth tends to reconcile models that predict similar suppression of the damping tail in the CMB temperature power spectrum, through the degeneracy $A_s e^{-2\tau_{\mathrm{reio}}}$. This result suggests that restricting the integration range and imposing specific reionization parameterizations in the standard approach may bias the inferred optical depth, which in that case becomes $\tau_{\mathrm{reio}} = 0.0531^{+0.0027}_{-0.0099}$ at 68\% CL~\cite{Planck:2018vyg}. For further discussion on the connection between the optical depth and the Hubble constant, see Refs.~\cite{Allali:2025wwi, Cain:2025usc}.

\begin{figure}[htbp]
    \centering
    \includegraphics[width=0.5\linewidth]{tau_reio.png}
    \caption{Normalized posterior distribution, $P/P_{\mathrm{max}}$, for the reionization optical depth, $\tau_{\mathrm{reio}}$, using the default definition implemented in the \texttt{CLASS} Boltzmann code. We find $\tau_{\mathrm{reio}} = 0.064^{+0.005}_{-0.026}$ at 68\% CL and $\tau_{\mathrm{reio}} = 0.064^{+0.061}_{-0.038}$ at 95\% CL. Results are derived from \textit{Planck} low-$\ell$ EE data alone.}
    \label{fig:tau_reio}
\end{figure}

\end{document}